\documentclass[journal=jacsat,manuscript=article]{achemso}

\usepackage[version=3]{mhchem} % Formula subscripts using \ce{}

\author{Chiwen Feng}
\affiliation{School of Physics and Optoelectronic Engineering, Guangdong University of Technology, Guangzhou 510006, China}

\author{Yanwei Liang}
\affiliation{School of Physics and Optoelectronic Engineering, Guangdong University of Technology, Guangzhou 510006, China}

\author{Jiaying Sun}
\affiliation{College of Chemistry, Zhengzhou University, Zhengzhou 450001, China}

\author{Renhai Wang}
\affiliation{School of Physics and Optoelectronic Engineering, Guangdong University of Technology, Guangzhou 510006, China}
\alsoaffiliation{Guangdong Provincial Key Laboratory of Sensing Physics and System Integration Applications, Guangdong University of Technology, Guangzhou 510006, China}
\email{wangrh@gdut.edu.cn}

\author{Huaijun Sun}
\affiliation{Jiyang College of Zhejiang Agriculture and Forestry University, Zhuji, 311800, China}
\email{hjsun@zafu.edu.cn}

\author{Huafeng Dong}
\affiliation{School of Physics and Optoelectronic Engineering, Guangdong University of Technology, Guangzhou 510006, China}
\alsoaffiliation{Guangdong Provincial Key Laboratory of Sensing Physics and System Integration Applications, Guangdong University of Technology, Guangzhou 510006, China}

\footnotetext[1]{C.F. and Y.L. contributed equally to this paper}
\title[An \textsf{achemso} demo]
  {  Predicting Miscibility in Binary Compounds:   
  A Machine Learning and Genetic Algorithm Study }

\abbreviations{IR,NMR,UV}

\DeclareUnicodeCharacter{2033}{''}
\begin{document}

\begin{tocentry}

\begin{center}
\includegraphics[width=\textwidth]{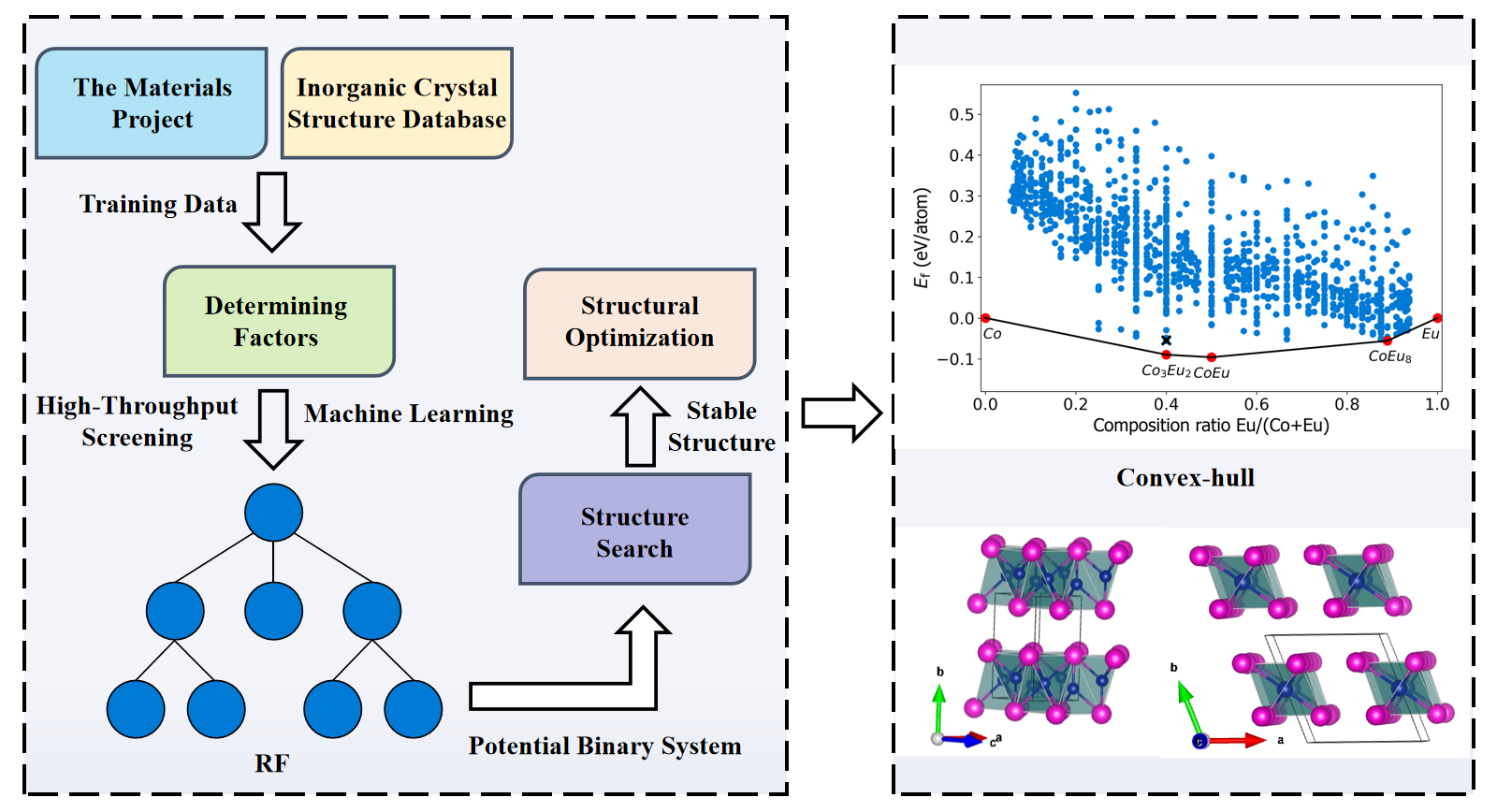}
\end{center}

\end{tocentry}


\begin{abstract}
The combination of data science and materials informatics has significantly propelled the advancement of multi-component compound synthesis research. This study employs atomic-level data to predict miscibility in binary compounds using machine learning, demonstrating the feasibility of such predictions. We have integrated experimental data from the Materials Project (MP) database and the Inorganic Crystal Structure Database (ICSD), covering 2,346 binary systems. We applied a random forest classification model to train the constructed dataset and analyze the key factors affecting the miscibility of binary systems and their significance while predicting binary systems with high synthetic potential. By employing advanced genetic algorithms on the Co-Eu system, we discovered three novel thermodynamically stable phases, \(CoEu_8\), \(Co_3Eu_2\), and \(CoEu\). This research offers valuable theoretical insights to guide experimental synthesis endeavors in binary and complex material systems.
\end{abstract}

\section{Introduction}
\begin{figure}
  \centering
  \includegraphics[scale=0.12]{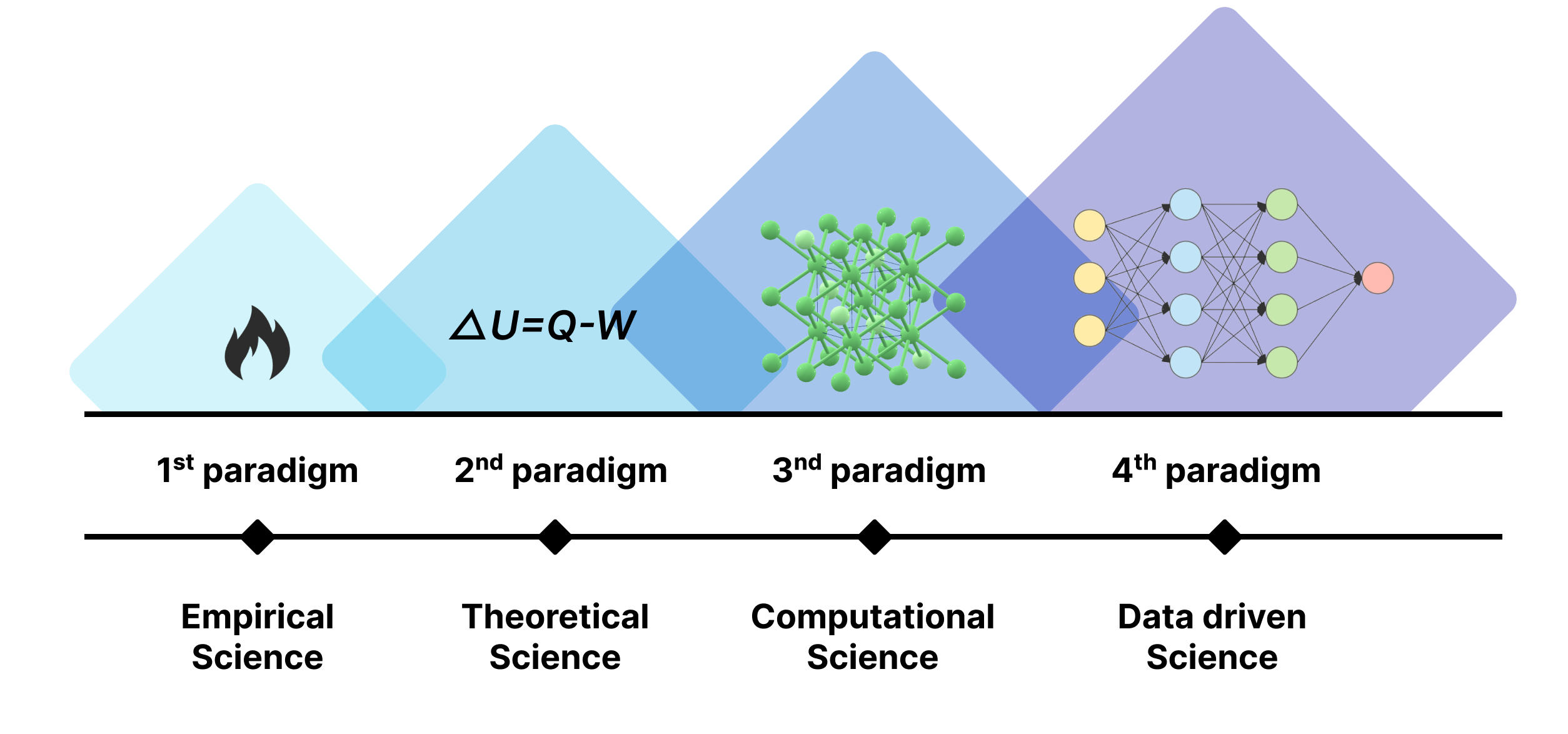}
  \caption{ Four paradigms of science: empirical, theoretical, computational and data-driven}
  \label{fgr:1}
\end{figure}

The field of materials science has consistently been a focal point for scientific research and innovation, leading to numerous in-depth and extensive R\&D activities. The evolution of materials science research has progressed through four major paradigms: experimental science, theoretical laws, computational modeling, and data-driven approaches (Figure~\ref{fgr:1}) \cite{agrawal2016perspective,medina2022accelerating}. The initial paradigm relied on experimental trial and error, a process characterized by long development cycles, potentially spanning 10-20 years and high experimental costs. 

Only a few centuries ago, materials science began to transition from trial-and-error methods to a more systematic and theoretical approach, spurred by the development of physical theoretical models and general laws such as thermodynamic constants. This shift marked a new focus on ``material design''. However, as calculations became more complex, theoretical computing in its second paradigm encountered significant bottlenecks in simulating intricate phenomena. It was not until the advancement of computers over the past decade that the third paradigm emerged, facilitating virtual laboratory simulations of real-world phenomena and enabling the ``synthesis'' of new materials.

First-principles calculations based on Density Functional Theory (DFT) \cite{lejaeghere2016reproducibility}, Local Density Approximation (LDA), and Generalized Gradient Approximation (GGA) \cite{perdew1996generalized} have been widely used to study the properties of new binary materials \cite{singh2023data,huang2024descriptor}. Meanwhile, significant advancements in ground-state structure prediction tools have completely transformed the field of materials science, making it possible to predict new material structures before experimental synthesis. For example, methods such as Genetic Algorithms/Evolutionary Algorithms \cite{chen2017improved,kadan2023accelerated}, Particle Swarm Optimization \cite{wang2012calypso,wang2010crystal}, Random Sampling \cite{pickard2011ab,pickard2009structures}, Minima Hopping \cite{goedecker2004minima,amsler2010crystal}, Simulated Annealing \cite{doll2007global}, Topological Modeling Method \cite{carlsson2020topological}, and Firefly Algorithm \cite{avendano2016firefly,yang2010firefly} have achieved notable success. Among them, is the Universal Structure Predictor: Evolutionary Xtallography (USPEX), known for its powerful search capabilities and high success rate \cite{oganov2006crystal,oganov2011evolutionary}. It performs well in finding low-energy structures of variable composition compounds, especially in the search for binary compounds \cite{ali2024prediction}.

However, given the vast unexplored space and the significant computational costs involved, traditional methods of exploration appear impractical. The rapid development of data mining and artificial intelligence technologies has propelled materials science into its fourth paradigm. Machine learning algorithms, known for their robust data processing capabilities and flexibility, can efficiently analyze large datasets without requiring preset hypotheses. Through machine learning, we can perform virtual high-throughput screening of extensive molecular spaces, rapidly identifying structures with potential\cite{barnard2014silico,nosengo2016material,curtarolo2013high}. This methodology not only expedites the discovery of novel materials but also significantly reduces both computational and experimental costs \cite{singh2023data,huang2024descriptor}.

The proposal of the Materials Genome Initiative \cite{hill2016materials} has significantly promoted the development of big data in materials science \cite{de2014materials}. A wealth of data has been accumulated through both experimental findings and computational simulations. The ICSD is currently the largest database for experimental identification of inorganic crystal structures, containing data dating back to 1913 and encompassing 291,382 fully evaluated and published crystal structure entries derived primarily from experimental results \cite{bergerhoff1983inorganic,belsky2002new}. The MP database includes 154,718 entries, providing researchers with comprehensive DFT data \cite{jain2013commentary,ong2013python}. The Open Quantum Materials Database (OQMD) is a high-throughput database that currently features nearly 300,000 compounds' DFT total energy calculations and common crystal structure information from ICSD \cite{saal2013materials,kirklin2015open}. These databases lay a solid foundation for the application of machine learning in materials science \cite{barnard2014silico,chen2022machine}.

Currently, there is no unified theoretical framework or clear concept to describe the key factors that determine whether two different elements can be miscible to form binary compounds. Therefore, we employed the machine learning method (random forest classifier) to train a model on the constructed dataset. The dataset consists of binary system data from experiments in MP database and ICSD, as well as data on immiscible binary systems. This method leverages the powerful features and flexibility of machine learning to address the classification problem of binary system miscibility. It identifies key factors influencing the miscibility of two different elements in forming binary compounds and analyzes the importance of these factors. Notably, this model predicts potentially miscible binary systems that have not yet been recorded in experimental databases. Using USPEX to quickly search for variable component structures of potentially miscible binary systems, and combining the Vienna Ab-initio Simulation Package (VASP) \cite{kresse1996efficiency} with high-precision optimization of the stable structures generated by USPEX, we successfully predicted the ground state structures of three different binary compounds. These structures demonstrate superior thermodynamic stability compared to theoretical data for the same systems in the OQMD, suggesting a greater advantage for experimental synthesis. Our theoretical research provides valuable guidance for discovering and synthesizing new compounds in similar systems.
\section{Computational Method}

\subsection{Machine Learning Model Building}

\subsubsection{Dataset}
The MP database does not include unstable elements like Po, At, Rn, Fr, Ra, Rf, Db, Sg, Bh, Hs, Mt, Ds, Rg, Cn, Nh, and Og, among others, which are either naturally radioactive or synthetically produced. Radioactive elements such as Ac, Th, Pa, U, Np, Pu, Tc, and Pm undergo spontaneous decay, leading to instability in the compounds they form. Noble gases —He, Ne, Ar, Kr, and Xe— have complete electron shells, making them chemically inert under normal conditions due to their high ionization energy. Halogens (F, Cl, Br, I) and elements like O, N, and H, known for their high electronegativity, multiple oxidation states, and reactivity, readily form a vast array of compounds.

Excluding these elements, our classification model will utilize a 2,346 possible binary system combinations dataset. We integrate experimental data from both the MP database and ICSD, encompassing 20,295 binary compounds across 1,420 binary systems. Binary systems that are experimentally miscible from these databases are used as positive samples, while the 926 non-miscible systems serve as negative samples. The balance of positive and negative samples in the data set enhances the model's performance and generalization capability.

\subsubsection{Determinants}
Drawing from experience, we explored several factors that could influence the synthesis of binary compounds and developed simple features, such as addition or subtraction, to represent their interrelationships, aiming to pinpoint key factors. Initially, we examined various parameters from the periodic table, including electronegativity, atomic radius, melting point, flash point, and molar volume. Notably, we observed significant differences in electronegativity between miscible and non-miscible binary systems, characterizing that electronegativity plays a key role in the miscibility of elements to form binary compounds. This observation led us to hypothesize that electronegativity is a pivotal factor in compound formation. Electronegativity quantifies an atom's capacity to attract electrons in a molecule, an element with a higher electronegativity value exerts a stronger pull on bonding electrons \cite{jensen1996electronegativity,cherkasov1998concept}. Consequently, we selected the sum of electronegativity values as a critical indicator for our study.

However, when comparing and analyzing other parameters, we did not observe the expected significant changes (see Figure S1), prompting the introduction of custom parameters. Empirically, the number of extranuclear electrons has been observed to play a significant role in the formation of binary compounds. However, there is currently no specific parameter available to quantify this effect. To address this, we propose a novel weighted indicator, termed WEle (Weighted Electrons of the Extranuclear Level), which quantifies the relationship between the extranuclear electron count of different elements and their contribution to compound stability. 

WEle is defined to more intuitively represent the stability of an atom or molecule's electron configuration in relation to its orbital capacity. A positive WEle value indicates an excess of electrons available for bonding, while a negative value suggests a need for interaction with other atoms to achieve stability. Consequently, we've chosen both the sum and difference of the valence electron weights as indicators, demonstrating the potential gain or loss of valence electrons for the elements involved in compound formation. The specific formula is shown below:
\begin{equation}
\text { WEle }=\frac{1}{0+1} * n_{s}+\sum \frac{1}{l+1} * n_{\mathrm{e}}
\begin{cases}
    n_{e}=-\left[2(2 \mathrm{l}+1)-n_{l}\right] & \text{if } n_{l}>2 l+1, \\
    n_{e}=n_{l} & \text{if } n_{l} \le 2 l+1.
\end{cases}
\end{equation}

In this formula, \(l\) represents the orbital type, with the `s', `p', `d', and `f' orbitals being assigned the numerical values 0, 1, 2, and 3, respectively. Additionally, \(n_{l}\) represents the current number of electrons occupying the orbital \(l\). When the outermost orbital of the extranuclear charge distribution is an `s' orbital, \(n_{s}\) specifically refers to the current number of electrons in that outermost s orbital, furthermore, \(n_{e}\) represents the ``effective" electron count in orbitals such as `p', `d', and `f', which are beyond the outermost shell.

We've also introduced a metric for the number of unoccupied orbitals for electrons, termed Unoccupy. Unoccupied orbitals indicate an atom's potential to achieve a stable electron configuration by sharing, transferring, or rearranging electrons, particularly within the outermost shell. The difference in the number of Unoccupy orbitals between elements is selected as an indicator, reflecting their potential for chemical bonding and stability.

Taking Fe as an example, its arrangement of extranuclear electrons is [(1, `s', 2), (2, `s', 2), (2, `p', 6), (3, `s', 2), (3, `p', 6), (3, `d', 6), (4, `s', 2)]. Based on the valence electron weight index formula, WEle can be calculated as \(\text { WEle }=\frac{1}{0+1} * 2+\frac{1}{1+2} * -4\), and the result is 0.667. Also, according to the definition of unoccupied orbitals, it’s determined that Unoccupy equals 4.

After careful consideration of various indicators, we have identified four critical parameters that significantly influence the miscibility of binary compounds: the sum of electronegativity values (sumElecNeg), the difference in the number of unoccupied orbitals (diffUnoccupy), the sum of valence electron weights (sumWEle), and the difference in valence electron weights (diffWEle), respectively.

\begin{equation}
\text { sumElecNeg }=ElecNeg_a + ElecNeg_b
\end{equation}
\begin{equation}
\text { sumWEle }=WEle_a + WEle_b
\end{equation}
\begin{equation}
\text { diffWEle }=|WEle_a - WEle_b|
\end{equation}
\begin{equation}
\text { diffUnoccupy }=|Unoccupy_a - Unoccupy_b|
\end{equation}

where \(a\) and \(b\) represent two different elements.

\subsubsection{Random Forest Model}
Our research focuses on exploring the key factors that determine whether two different elements can be miscible to form binary compounds and quantifying the importance of these factors. There are the following characteristics in the data of this study: (1) high data dimension, involving the basic characteristics of many elements; (2) large data scale, with a large number of samples to be processed; (3) intricate relationship between the data, preceding a simple linear relationship. Given these characteristics, the random forest model is an ideal choice due to its exceptional fitting capability and its proficiency in managing complex data structures \cite{breiman2001random,chandy2023mim}.

We utilized data from 2,346 binary systems to construct our random forest model. To effectively evaluate the model's performance, we employed a commonly used partitioning strategy, randomly dividing the dataset into a training set and a test set with an 80\% to 20\% ratio. This method ensures that the model does not overfit during training while allowing for effective performance evaluation on an independent dataset.

During the training process, we use information entropy as a criterion to evaluate the quality of node splitting because it can measure the data's uncertainty and evaluate each node's purity after splitting \cite{shannon1948mathematical}. By employing this method, we can identify the optimal split points, thereby enhancing the model's performance to its fullest potential. We integrated a subset of 10 decision trees into the random forest model. Through this integration strategy, we enhanced the prediction accuracy and robustness of the model, thereby effectively improving its generalization ability.

To comprehensively evaluate the performance of the constructed random forest model in classifying the miscibility of binary compound systems, we utilized stratified sampling to randomly generate the test set. Additionally, we assessed the model using four key performance metrics: Accuracy, Precision, Recall, and F1-score. The calculation formulas for these metrics are as follows:

Accuracy measures the proportion of correctly classified samples to the total number of samples:
\begin{equation}
Accuracy =\frac{TP+TN}{TP+TN+FP+FN} 
\end{equation}

Precision measures the proportion of true positive samples among all samples classified as positive:
\begin{equation}
Precision =\frac{TP}{TP+FP} 
\end{equation}

Recall measures the proportion of true positive samples correctly identified among all actual positive samples:
\begin{equation}
Recall =\frac{TP}{TP+FN} 
\end{equation}

F1-score is the harmonic mean of precision and recall, providing a comprehensive metric that balances both aspects:
\begin{equation}
F1=\frac{2Precision*Recall}{Precision+Recall} 
\end{equation}

Here, \(TP\), \(TN\), \(FP\), and \(FN\) represent true positives, true negatives, false positives, and false negatives, respectively. This evaluation framework ensures a thorough analysis of the model’s performance across multiple dimensions.

The key performance metrics we used, such as accuracy, precision, recall, and F1-score, were calculated using weighted averages based on class weights. These weights were assigned according to the size of each class to ensure balanced predictive performance across all classes. We performed numerous experiments with random combinations of potential tuning parameters to identify the optimal set of hyperparameters. Finally, we applied the best-fitting random forest model to the test set to evaluate its performance on unseen data.

\subsection{Structural Prediction}
Recognized for its efficiency in identifying all stable crystal structures across a range of atomic numbers, USPEX has been validated through numerous studies \cite{oganov2004theoretical,ma2009transparent,zhang2013unexpected}. We employed the USPEX algorithm to conduct a structural search at 0 GPa. The search began with an initial population of 150 diverse structures, each with up to 16 atoms in their respective primitive cells. The evolutionary process involved generating subsequent populations of 40 structures. This generation was composed through a combination of methods: 20\% heredity, ensuring the propagation of successful structures; 20\% random space group generation, introducing new symmetry possibilities; 20\% transmutation, simulating atomic substitutions; 20\% soft mutation, allowing for slight modifications to existing structures; 10\% lattice mutation, altering the unit cell parameters; and 10\% topological random generation, exploring new topological configurations.

The search algorithm is designed to terminate under two conditions: either when the optimal structure has remained unchanged for 30 consecutive generations, indicating convergence towards a stable solution, or after completing 60 iterations, ensuring a thorough exploration of the configurational space.

We employed the VASP code to carry out DFT calculations using the Perdew–Burke–Ernzerhof (PBE) functional within the framework of the GGA. To accurately represent the core electrons and their effects on the valence electrons, we utilized the projector-augmented wave (PAW) method \cite{sui2009influence}. 

A plane-wave basis was used with a kinetic energy cutoff of 500 eV, and the convergence criterion for the total energy was set to $10^{-8}$ eV. The unit cell lattice vectors (both unit cell shape and size) were fully relaxed together with the atomic coordinates until the force on each atom was less than 0.005 eV/Å.

To characterize the energetic stability of structures at various compositions resulting from a structure search, the formation energy (\(E_f\)) of any given structure \(E(Co_xEu_y)\) is calculated:
\begin{equation}
E_f=\frac{E(Co_xEu_y)-xE(Co)-yE(Eu)}{x+y} 
\end{equation}

Where \(E(Co_xEu_y)\) is the per-atom energy of \(Co_xEu_y\) structure, \(E(Co)\) and \(E(Eu)\) represents the per-atom energy of Hexagonal-Co (\(Fm\)-\(3m\)) and Hexagonal-Eu (\(P6_3/mmc\)), respectively (all details are provided in Table S3 of the Supporting information). Also, the phonon calculations were performed using the Density-Functional Perturbation Theory (DFPT) by combining the VASP code and PHONOPY package \cite{togo2008first,togo2015first}. A 2 × 2 × 2 supercell and a kinetic energy cutoff of 500 eV were used to calculate the phonon spectra.

\section{Results And Discussion}

\subsection{Machine Learning Prediction Model}

\begin{figure}
  \centering
  \includegraphics[scale=0.55]{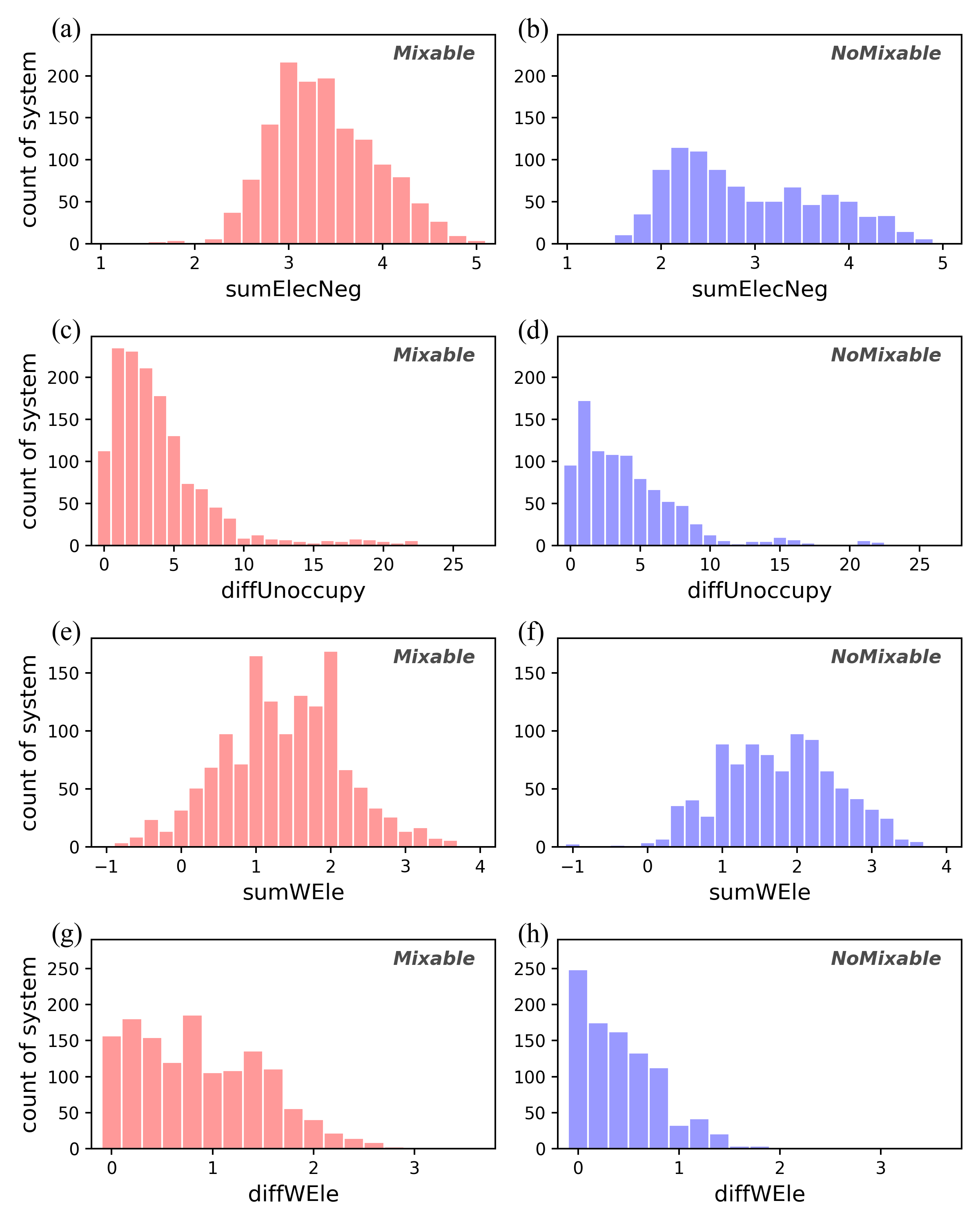}
  \caption{ Distribution of the selected classification labels in the model across binary systems }
  \label{fgr:2}
\end{figure}

Here, we have selected four critical factors: sumElecNeg, diffUnoccupy, sumWEle, and diffWEle. Figure~\ref{fgr:2} illustrates the relationships between the distribution of these key factors and the miscibility of different elements to form binary systems. The horizontal axis represents the four key factors, while the vertical axis represents the number of binary systems.

As seen from Figure~\ref{fgr:2}, a, c, e, and g represent the sample label distributions that can form binary compound systems, while b, d, f, and h correspond to the sample label distributions for which no binary compound systems have been found yet.

The effect of sumElecNeg on the binding of binary compounds is shown in Figure~\ref{fgr:2}(a-b). In Figure~\ref{fgr:2}(a), mixable phases are cumulative in the range of 2.8-4. On the contrary, nonmixable ones range between 2 and 3 and fail to get binary phases. What’s more, Figure~\ref{fgr:2}(c-d) reflects the effects of diffUnoccupy on the binding of binary compounds. In Figure~\ref{fgr:2}(c), mixable compounds decay sharply in the region after 2, while in Figure~\ref{fgr:2}(d), nonmixable compounds decay slowly in the region after 2. The effects of sumWEle on the binding of binary compounds are shown in Figure~\ref{fgr:2}(e-f). In Figure~\ref{fgr:2}(e), mixable phases are mainly concentrated in the region of 1-2, while in Figure~\ref{fgr:2}(f), nonmixable phases are mostly concentrated in the region of 1-2.5. Figure~\ref{fgr:2}(g-h) reflects the effects of diffWEle on the binding of binary compounds. In Figure~\ref{fgr:2}(g), mixable compounds mainly decay in the region after 1.5, while in Figure~\ref{fgr:2}(h), nonmixable drops in the region after 0.

We selected the above four predominant factors during the training of the Random Forest algorithm. Performance indicators were applied to test the model's performance. The specific indicators are as follows: Accuracy is 94.47\%, Precision is 94.83\%, Recall is 94.55\%, and F1-score is 94.64\%. The complete specific performance parameters are shown in Table~\ref{tbl:1}(Partial classification results for different labels in the binary system are shown in Table S2 in the supporting material).

\begin{table}
  \caption{Binary system classification results}
  \label{tbl:1}
  \begin{tabular}{lllll}
    \hline
    Mix or No-Mix & Precision & Recall & F1-Score & Support  \\
    \hline
    0 & 0.91 & 0.94 & 0.92 & 170  \\
    1 & 0.97 & 0.95 & 0.96 & 300  \\
    accuracy & ~ & ~ & 0.94 & 470  \\
    weighted avg & 0.95 & 0.94 & 0.94 & 470 \\
    \hline
  \end{tabular}
\end{table}

To further analyze the importance of each input, we extracted the feature contributions from the random forest model. As depicted in Figure~\ref{fgr:3} below, their weight ratios are 35.2\% (sumElecNeg), 22.54\% (diffWEle), 28.81\% (sumWEle), and 13.45\% (diffUnoccupy), respectively. SumElecNeg emerges as the most significant factor, highlighting the critical role of electronegativity summation in predicting the formation of binary compounds. Then it comes to sumWEle and diffWEle, which shows that the WEle weight value plays an important role in forming the compound. Also, diffUnoccupy provides useful global information about the formation propensity of a compound. These varying degrees of significance not only elucidate the extent to which these critical factors influence the formation of binary compounds by elements but also represent the degree to which distributions are differentiated in graphical analyses.

\begin{figure}
  \centering
  \includegraphics[scale=0.3]{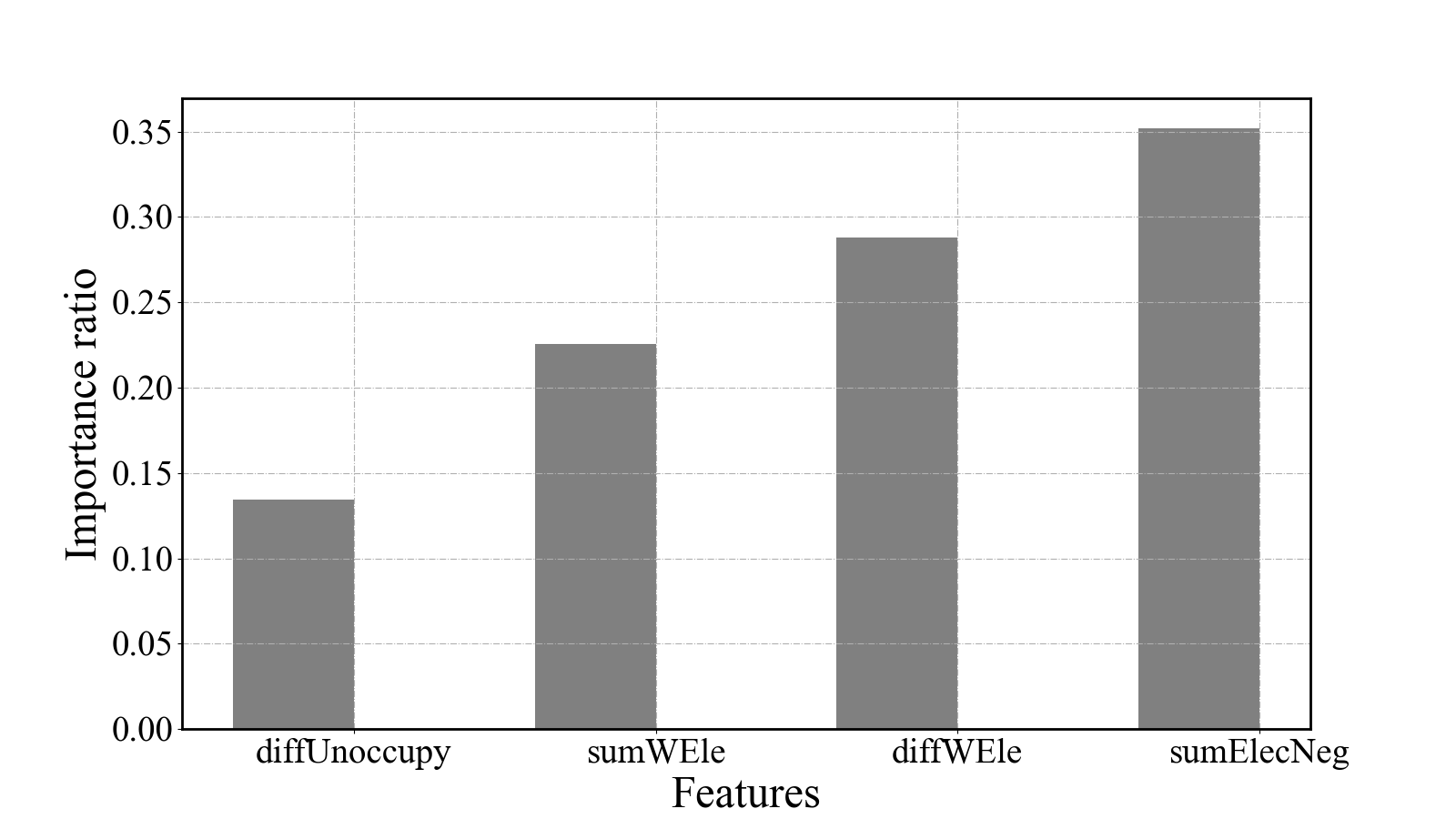}
  \caption{ The feature importance of random forest model }
  \label{fgr:3}
\end{figure}

Finally, the trained model has been applied to predict, some potential combinations that have the potential to form binary compounds but have not yet been found in the experimental database, including Na-Fe, Al-Bi, P-Te, Fe-Rb, Co-Eu, As-Tl, Mo-Tb, Ru-Pb, Ru-Bi, Pd-Os, Tb-W, and Ho-W. These predicted binary combinations will serve as our target set for discovering new materials.

\subsection{Updating The Energy Convex Hull Of Co-Eu System}

\begin{figure}
  \centering
  \includegraphics[scale=0.5]{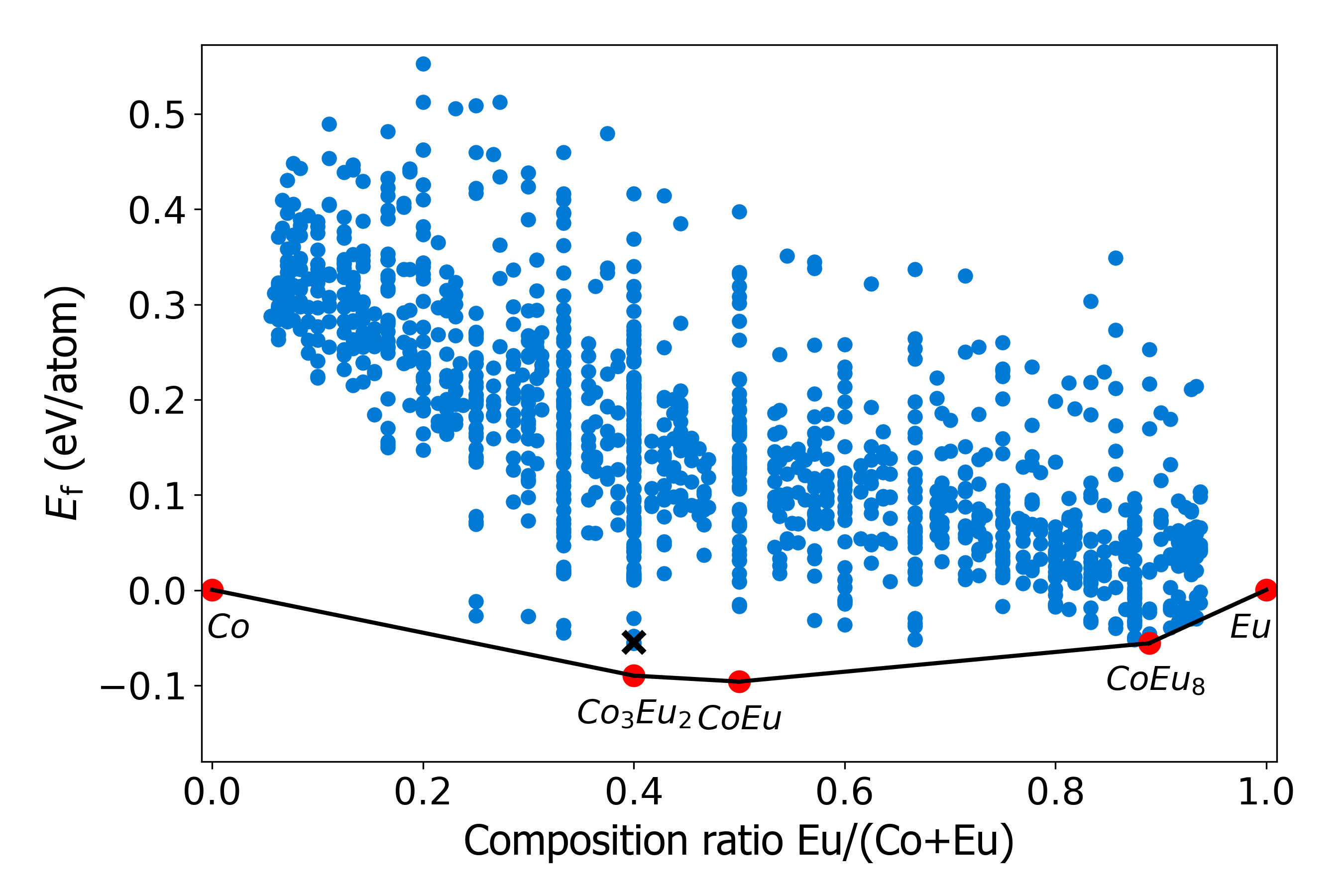}
 \caption{ This figure illustrates the evolution of the energy convex hull during the USPEX search. The solid black line represents the energy convex hull for the updated Co-Eu system. The `x' symbols indicate the current stable phases. }
  \label{fgr:4}
\end{figure}

We conducted a rapid variable component structure search using USPEX, guided by the results from the Random Forest Model Prediction Combination previously mentioned. It's known that the energy convex hull (ECH) phase diagram is a vital bias for judging the structural stability of a special system\cite{ali2024prediction}. In Figure~\ref{fgr:4}, red points represent stable phases, connected by continuous black lines to form the ECH. Phases directly on the ECH are thermally stable, while blue points above the hull indicate instability. Notably, the only currently stable phase, \(Eu_2Co_3\), was identified in the OQMD \cite{saal2013materials,kirklin2015open}. However, this phase has become metastable following the update of the ECH, and its formation energy is now marked as an `x' in the phase diagram. As shown in Figure~\ref{fgr:4}, \(CoEu_8\),  \(Co_3Eu_2\) and \(CoEu\) exhibit negative formation energy, indicating they are thermodynamically stable.

The structures of the lowest energy \(CoEu_8\),  \(Co_3Eu_2\) and \(CoEu\) are both in the \(P1\) space group, as shown in Figure~\ref{fgr:5}. Crystallographic details are provided in the supplementary materials (Tables S4-S6).

\begin{figure}
  \centering
  \includegraphics[scale=0.4]{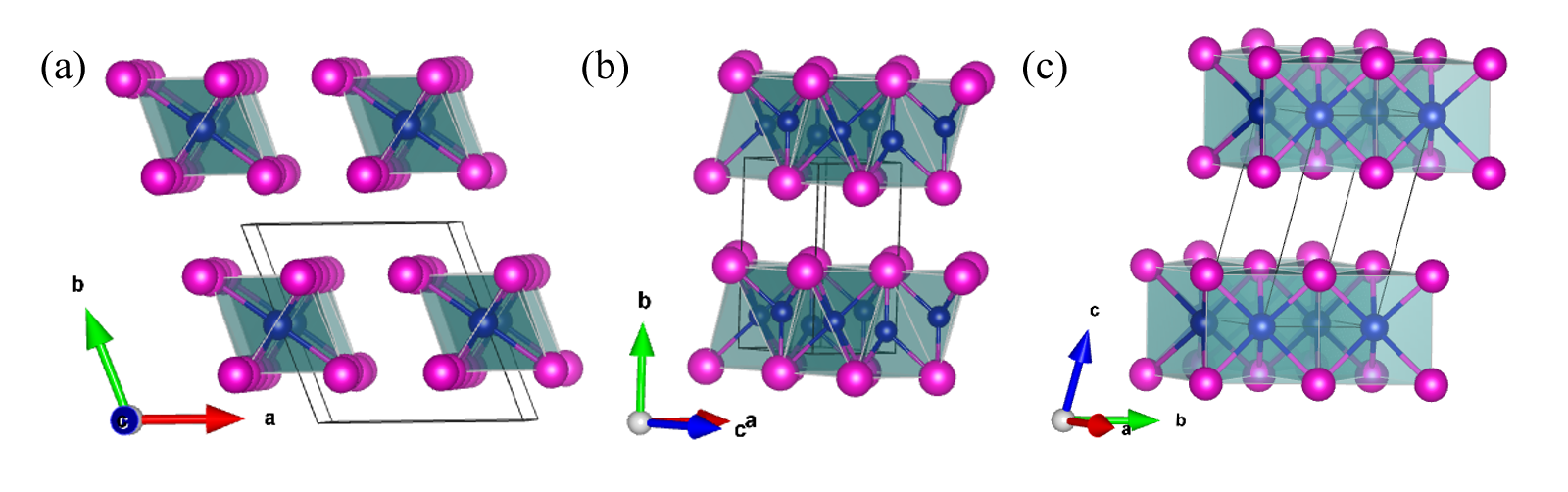}
  \caption{ Crystal structures of \(CoEu_8\)(a), \(Co_3Eu_2\)(b) and \(CoEu\)(c) phases. Eu atoms (the largest spheres) are colored in red, and Co atoms (the smallest spheres) are colored in blue. }
  \label{fgr:5}
\end{figure}

\begin{figure}
  \centering
  \includegraphics[scale=0.28]{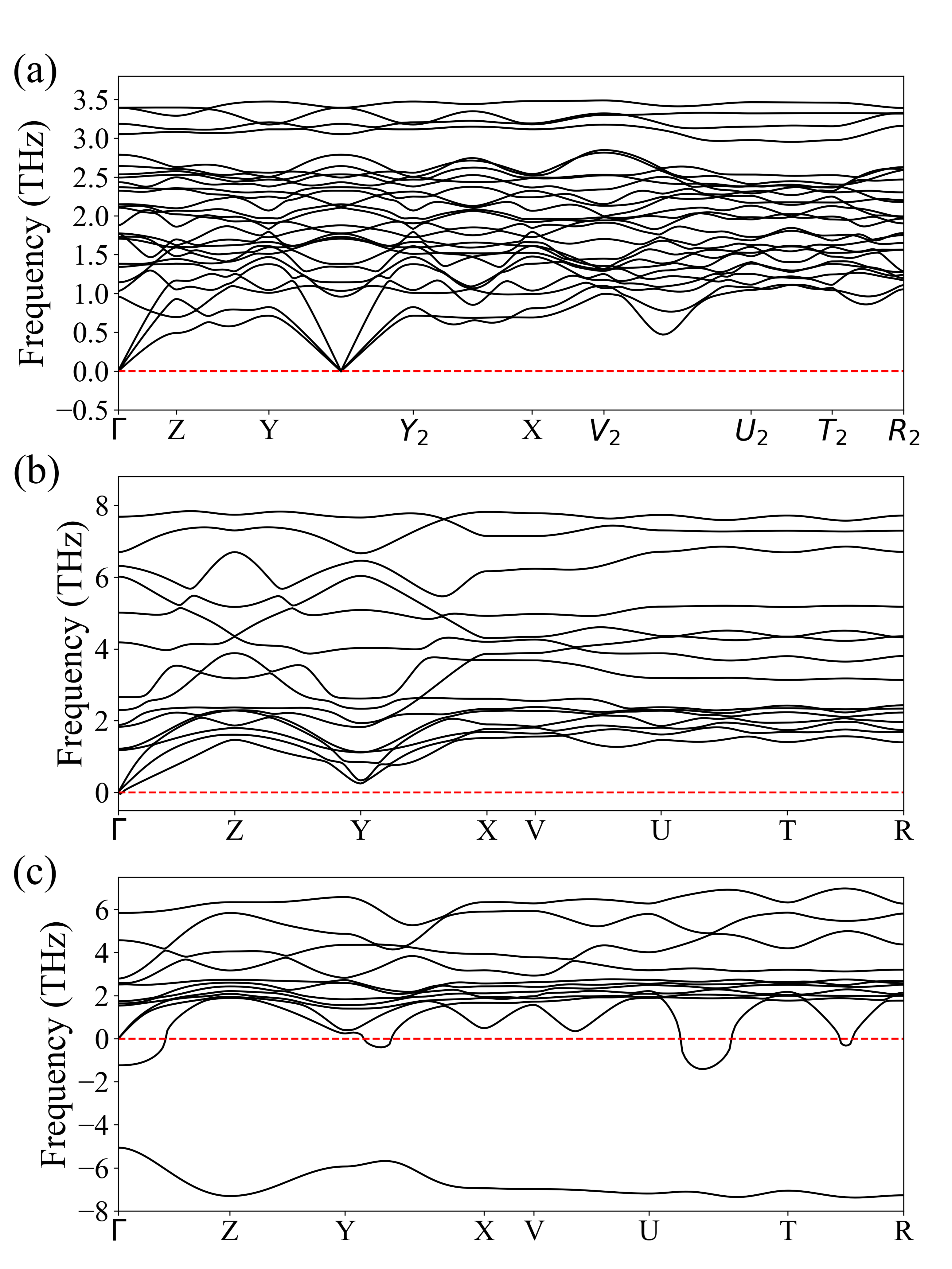}
  \caption{ Phonon band structures of \(CoEu_8\)(a), \(Co_3Eu_2\)(b) and \(CoEu\)(c). }
  \label{fgr:6}
\end{figure}

The phonon band structure offers key insights into the dynamic stability of newly predicted phases of \(CoEu_8\), \(Co_3Eu_2\) and \(CoEu\). The \(CoEu_8\) and \(Co_3Eu_2\) phases are dynamically stable, as there are no imaginary frequencies in Figure~\ref{fgr:6}(a-b), which further consolidates the theoretical basis for \(CoEu_8\) and \(Co_3Eu_2\) as potentially stable phases. Among them, since the \(CoEu\) phase has an imaginary frequency in Figure~\ref{fgr:6}(c), it indicates that it may be dynamically unstable. The phonon spectrum of \(Eu_2Co_3\) in OQMD after structural optimization also has imaginary frequencies, as shown in the supplementary material (Figure S7). 

Our results indicate that the newly predicted structures of the Co-Eu system exhibit superior stability, making them strong candidates for experimental synthesis. This finding corroborates the results presented in Chapter 3.1.

\section{Conclusion}

In summary, our study utilized a dataset comprising 2,346 binary systems, including 1,420 experimentally synthesized binary systems from the MP database and ICSD as positive samples and 926 experimentally unsynthesized binary systems as negative samples. We employed a random forest model for binary classification, integrating key features such as the sumElecNeg for two different elements in the periodic table and custom-weighted features like sumWEle, diffWEle, and diffUnoccupy. This feature set significantly bolstered the model's reliability, yielding an impressive Accuracy of 94.47\%, Precision of 94.83\%, Recall of 94.55\%, and an F1-score of 94.64\%.

We utilized USPEX to conduct rapid variable composition structure searches on 12 systems with synthetic potential predicted by our model and discovered three new thermodynamically stable structures, \(CoEu_8\), \(Co_3Eu_2\) and \(CoEu\), in the Co-Eu system, thereby verifying the reliability of our model. Our model successfully explored combinations of miscible element pairs, allowing us to explore and synthesize potentially stable structures more efficiently, providing valuable guidance for future discovery of new materials. It also laid the foundation for further investigations into incorporating a third element to form new ternary systems with previously immiscible element pairs.

\begin{acknowledgement}

Work at Guangdong University of Technology is supported by the Guangdong Basic and Applied Basic Research Foundation (Grant No. 2022A1515012174 and 2021A1515110328). R. Wang also thanks Center of Campus Network and Modern Educational Technology, Guangdong University of Technology, Guangdong, China for providing computational resources and technical support for this work.

\end{acknowledgement}

\begin{suppinfo}

The Supporting Information can be accessed at no cost through https://pubs.acs.org/doi/xxx.

\end{suppinfo}

\bibliography{achemso-demo}

\end{document}